\documentclass[conference]{IEEEtran}
\IEEEoverridecommandlockouts

\usepackage{cite}
\usepackage{url}
\usepackage{amsmath,amssymb,amsfonts}
\usepackage{algorithmic}
\usepackage{graphicx}
\usepackage{booktabs}
\usepackage{balance}
\usepackage{textcomp}
\usepackage[dvipsnames]{xcolor}
\usepackage{subcaption} 
\usepackage{hyphenat}
\usepackage{amsmath}
\usepackage{amsfonts}
\usepackage{amssymb}
\DeclareRobustCommand{\rub}[1]{\textcolor{black}{#1}} 
\DeclareRobustCommand{\greenub}[1]{\textcolor{black}{#1}} 
\usepackage[colorinlistoftodos]{todonotes}
\usepackage{makecell}
\def\BibTeX{{\rm B\kern-.05em{\sc i\kern-.025em b}\kern-.08em
    T\kern-.1667em\lower.7ex\hbox{E}\kern-.125emX}}

\usepackage{soul}
\usepackage{multirow}
\usepackage{booktabs}
\usepackage{tabularx}
\usepackage{tikz}
\usepackage{tikzpagenodes}
\usetikzlibrary{calc}

\AddToHook{shipout/foreground}{%
    \ifnum\value{page}=1
        \begin{tikzpicture}[remember picture, overlay]
            \node[
                draw,
                text width=0.95\textwidth,
                align=left,
                font=\footnotesize,
                inner sep=5pt
            ]
            at ($(current page header area.center) + (0,-5pt)$)
            {%
                This paper has been accepted for publication in the proceedings
                of the IEEE International Conference on Sensing, Communication, and Networking (SECON), 2026. This is the authors' accepted version
                of the article. The final version published by IEEE is:
                T.~Jiang, A.~Alchaab, A.~Younis and D.~Pompili, ``CoMeT-Net: Consensus Memory Template Network
               for Real-time Traffic Anomaly Detection,'' in \textit{Proc. of
                the IEEE International Conference on Sensing, Communication, and Networking (SECON)}, Pisa, Italy, June 2026.
            };
        \end{tikzpicture}%
    \fi
}

\begin{document}

\title{CoMeT-Net: Consensus Memory Template Network for Real-time Traffic Anomaly Detection}
\author{{\bf Tingcong Jiang*†, Adhwaa Alchaab*†, Ayman Younis, and Dario Pompili†}\\
†Department of Electrical and Computer Engineering, Rutgers University--New Brunswick, NJ, USA\\
\textit{\{tingcong.jiang, adhwaa.alchaab , a.younis, pompili\}@rutgers.edu}
\thanks{* Tingcong Jiang and Adhwaa Alchaab contributed equally to this work.}
}

\maketitle
\begin{abstract}
Real-time anomaly detection in Open Radio Access Networks~(O-RAN) demands high accuracy, low false alarms, and computational efficiency for resource-constrained edge deployment. Traditional methods struggle with computational overhead, inconsistent cross-domain performance, and suboptimal feature representations that miss subtle attacks on O-RAN's open interfaces. We present CoMeT-Net (Consensus Memory Template Network), a framework achieving state-of-the-art detection through three innovations: (1)~structured memory banks enabling template-based consensus voting with $O(N \cdot C)$ complexity; (2)~adaptive gating that downweights ambiguous features as a learned noise filter; (3)~contrastive alignment unifying feature learning and classification. Deployed in O-RAN infrastructure via edge servers and Near-RT RIC xApp, CoMeT-Net enables dynamic threat mitigation through PRB throttling and RRC connection release. On network traffic datasets, CoMeT-Net achieves 99.35\% F1 score with 10$\times$ lower false alarm rates than baselines while maintaining 0.3-3ms inference across hardware tiers from servers to Raspberry Pi 4. O-RAN testbed validation demonstrates effective isolation, degrading attacker latency to $>$1400ms while preserving 15-20ms for legitimate users.


\end{abstract}

\section{Introduction}

Open Radio Access Network~(O-RAN) security faces critical challenges from cyber threats that exploit open interfaces and distributed architectures. Consider a Distributed
Denial-of-Service~(DDoS) attack: malicious users flood the network with traffic, rapidly consuming Physical Resource Blocks~(PRBs), which are the fundamental radio resources allocated to users, and degrading service for legitimate users.
Without real-time detection and mitigation, such attacks quickly overwhelm the infrastructure, violating Service Level Agreements~(SLAs) and causing network outage. Intrusion Detection Systems~(IDS) provide real-time information for applications in the Near-Real-Time RAN Intelligent Controller (Near-RT RIC) to trigger countermeasures including dynamic Physical Resource Block~(PRB) allocation and Radio Resource Control~(RRC) management against attacks ranging from  DDoS to Advanced Persistent Threats~(APTs). O-RAN's disaggregated architecture enables distributed anomaly detection through edge servers co-located with User Plane Functions~(UPF) for proximal threat detection and xApps orchestrating network-wide responses via standardized E2 interfaces with gNodeBs~(gNBs). This enables adaptive resource throttling for malicious users while preserving Quality of Service~(QoS) across network slices.

\noindent\rub{O-RAN's time-critical requirements create unique challenges.} Unlike \rub{anomaly detection in} financial fraud \greenub{or} healthcare monitoring~\cite{pinto2022literature}, where seconds to minutes of latency may be acceptable.
O-RAN \rub{demands subsecond detection}~\cite{huang2025meta}. \rub{A delayed response to an adversarial event could allow malicious traffic to monopolize PRBs, triggering service degradation across slices.} \noindent\rub{An effective O-RAN anomaly detection framework must balance four critical requirements:} \greenub{(1)~\emph{computational efficiency}, sub-millisecond inference to enable real-time detection before PRB exhaustion;} \greenub{(2)~\emph{adaptability}, recognizing evolving attack patterns beyond training data;} \greenub{(3)~\emph{interpretability}, providing clear evidence to justify resource throttling decisions for regulatory compliance; and} \greenub{(4)~\emph{integration}, native compatibility with E2 Service Models and RIC platforms for automated mitigation.} \rub{These requirements are essential} \greenub{because} dynamic resource allocation \greenub{decisions (such as reducing an attacker's PRB quota from 50\% to 5\%)} directly impact\greenub{s} QoS across \greenub{all} network slices\greenub{, from enhanced mobile broadband to ultra-reliable low-latency communications}.


\begin{figure}[t]
\centering
\includegraphics[width=0.5\textwidth]
{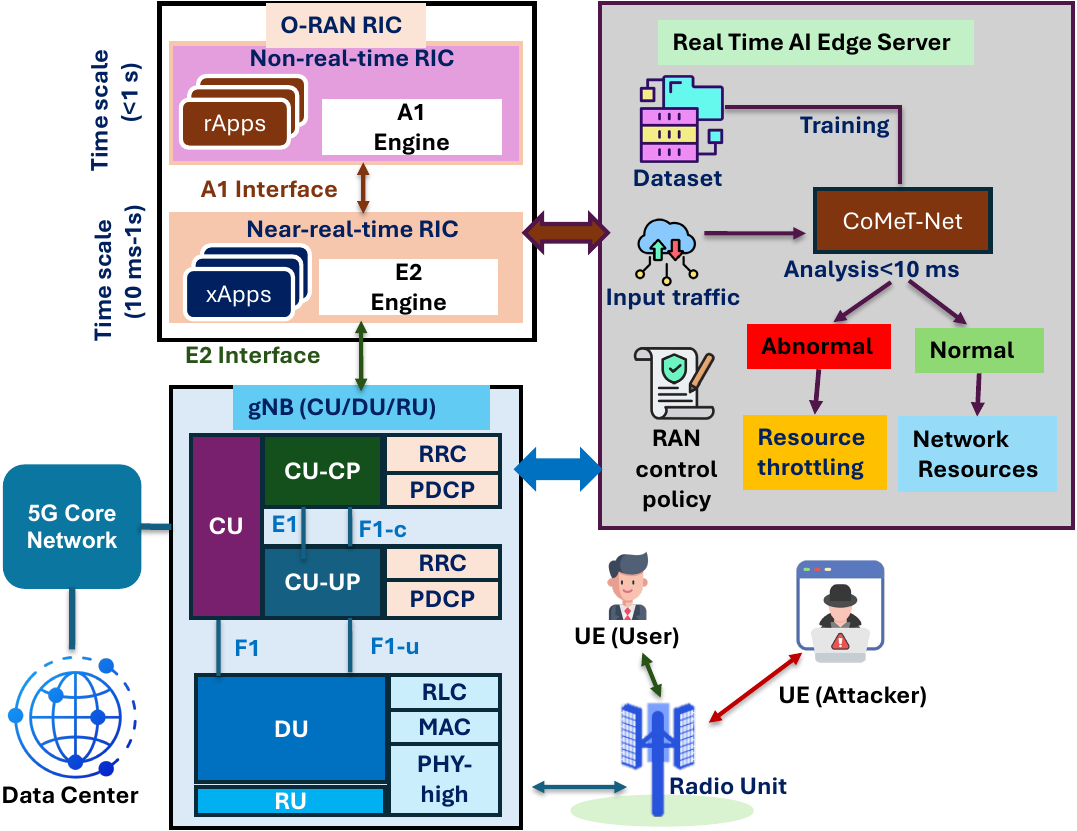}
\vspace{-0.1in}
\caption{CoMeT-Net: Closed-loop, anomaly-driven dynamic PRB reallocation in O-RAN, with real-time detection at the AI edge and mitigation via near-RT RIC control.}
\label{fig:NET}
\vspace{-0.3in}
\end{figure}
\noindent\textbf{Motivation:}
\rub{Existing approaches face critical limitations}\greenub{:} \emph{\rub{Reconstruction methods}}~\cite{torabi2023practical, iqbal2023reconstruction, sabuhi2021applications} require seconds and many resources to detect anomalies, allowing adversarial attacks to exhaust PRBs before mitigation. \emph{\rub{Ensemble methods}}~\cite{breiman2001random, Chen_2016} show inconsistent performance and struggle to deploy on resource-constrained edge servers. \emph{\rub{Classification methods}} are computationally efficient for O-RAN's latency requirements, but they often learn weak traffic representations in disaggregated RAN pipelines. Because supervision is typically task-level rather than feature-level, the resulting embeddings may not reliably separate benign traffic from attacks, missing subtle cross-slice patterns, leading to false alarms and misallocation of resources. Moreover, dense representations lack interpretability (critical for regulatory compliance), and existing systems lack native E2 integration, requiring manual intervention rather than automated mitigation.

\noindent\textbf{Our Approach:} \rub{To address these challenges}\greenub{, we propose} CoMeT-Net (Consensus Memory Template Network)\greenub{, which integrates detection and mitigation in O-RAN. As shown in Fig.~\ref{fig:NET}, CoMeT-Edge servers co-located with UPFs analyze traffic in real time, while a CoMeT-xApp in the Near-RT RIC coordinates automated responses via E2 to throttle PRB allocation and release RRC connections for malicious users.}

\greenub{CoMeT-Net introduces three innovations for anomaly detection:} \emph{\rub{(1)~Structured memory bank}}\greenub{:} \rub{stores learned templates} \greenub{(e.g., normal vs. attack traffic patterns). Each traffic sample generates multiple feature vectors that vote by comparing against templates, eliminating classifier networks for sub-millisecond inference and providing interpretable decisions.} \emph{\rub{(2)~Adaptive gating}}\greenub{:} \rub{weighs vote importance}\greenub{, allowing features to abstain when ambiguous. For instance, if packet size distributions are similar between DDoS and legitimate traffic, that feature contributes less, focusing on discriminative signals like flow rates.} \emph{\rub{(3)~Contrastive alignment loss}}\greenub{:} \rub{simultaneously optimizes} \greenub{feature learning and classification, pushing normal and DDoS traffic representations apart to form distinct clusters, enabling detection of subtle attack variations.}


\rub{\textbf{Our Contributions} are}\greenub{:}
\begin{itemize}
\item \rub{We propose CoMet-Net }\greenub{with template-based consensus voting and adaptive gating, achieving 99.35\% F1 score (10$\times$ lower false alarms) on network intrusion detection with 0.3-3~ms inference across server-to-IoT platforms.}
\item \greenub{Contrastive alignment loss that jointly optimizes feature learning and classification, creating distinct normal/attack clusters for interpretable decisions.}
\item \greenub{End-to-end O-RAN integration: CoMeT-Edge servers for real-time detection and CoMeT-xApp for automated E2-driven PRB throttling and RRC release.}
\item \greenub{Testbed validation demonstrating effective attacker isolation while preserving legitimate user QoS.}
\end{itemize}

\noindent\textbf{Paper Outline:} \greenub{Sect.~\ref{sec:related_work} surveys related work. Sect.~\ref{sec:proposed_work} presents CoMeT-Net: memory bank, gating, and contrastive alignment. Sect.~\ref{sec:performance_eval} evaluates performance across datasets and O-RAN testbed. Sect.~\ref{sec:future_work} summarizes this manuscript and discusses future directions.}

\section{Related Work} \label{sec:related_work}
\noindent\textbf{Network Anomaly Detection:} \greenub{Traditional approaches rely on statistical methods and classical classifiers. Na\"ive Bayes and SVMs~\cite{708428, scholkopf2001estimating} struggle with complex attack patterns due to predefined kernel functions and poor scaling with dataset size. Ensemble methods emerged as stronger alternatives: Random Forests~\cite{breiman2001random} and Gradient Boosting implementations (XGBoost~\cite{Chen_2016}, LightGBM~\cite{ke2017lightgbm}, CatBoost~\cite{prokhorenkova2018catboost}) achieve strong performance on tabular data but lack interpretability and optimal feature learning for network traffic.}

\greenub{For O-RAN deployment,~\cite{reis2025edge} presents federated learning achieving F1$>$0.91 with sub-20~ms latency, though lacking feature-level interpretability.~\cite{kakani2024evaluation} establish the O-RAN security landscape and identify vulnerabilities.~\cite{awad2024xapps} demonstrates ML-driven xApps for V2X scenarios. However, these solutions employ separate detection and classification modules, increasing computational overhead and reducing interpretability.}

\begin{figure*}[ht!]
    \centering
    \vspace{-0.3cm}
    \includegraphics[width=0.8\linewidth]{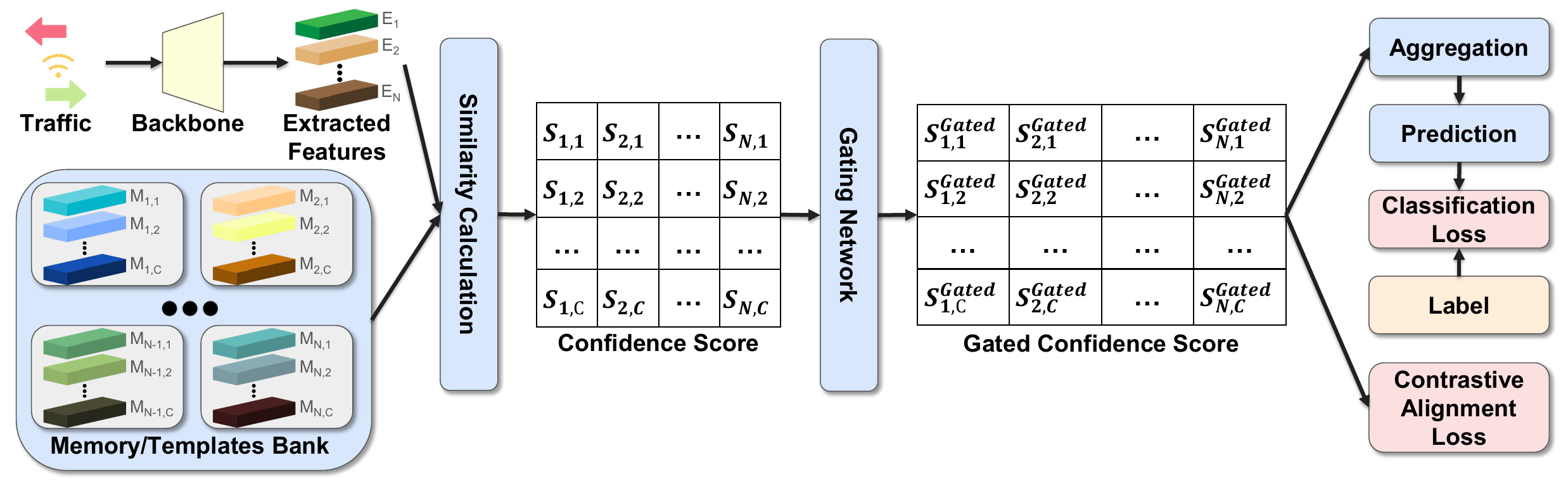}
    \caption{The proposed CoMeT-Net architecture showcasing our novel memory-based consensus voting approach. Our key innovations include: (1) a structured memory bank for template-based knowledge representation, (2) a direct comparison mechanism and gating mechanism that replaces classifiers to reduce overhead, and (3) a unified optimization approach that aligns feature learning with classification objectives, enabling state-of-the-art performance across diverse domains.}
    \vspace{-0.3cm}
    \label{fig:enter-label}
\end{figure*}

\noindent\textbf{Deep Learning Approaches:} \greenub{Reconstruction-based methods~\cite{ ye2024novel, iqbal2023reconstruction} model normal data distributions and flag deviations as anomalies. While capable of detecting novel attacks, these impose substantial computational overhead during training and inference, hindering real-time deployment. Memory-augmented autoencoders (MemAE)~\cite{gong2019memorizing} use memory modules for reconstruction but require computationally intensive compression, matching, and reconstruction steps, making them unsuitable for O-RAN's sub-millisecond latency requirements.}

\greenub{Classification-based approaches~\cite{iqbal2024anomaly, hairab2022anomaly} offer faster inference but struggle with feature representation learning. Current optimization objectives supervise only the final classification without explicitly guiding discriminative feature learning, resulting in suboptimal feature spaces that miss subtle attack patterns. Transformer models like FTTransformer~\cite{gorishniy2021revisiting} face quadratic complexity ($O(N^2)$) that limits real-time deployment, and produce dense representations that lack interpretability, critical for regulatory compliance in security contexts.}


\begin{table}[t]
\centering
\caption{\greenub{CoMeT-Net vs. existing approaches.}}
\label{tab:comparison}
\small
\begin{tabular}{@{}lccc@{}}
\toprule
\textbf{Approach} & \textbf{Complexity} & \textbf{Interpretable} & \textbf{E2 Native} \\ \midrule
MemAE~\cite{gong2019memorizing} & $O(N^2)$ (recon.) & No & No \\
Transformers~\cite{gorishniy2021revisiting} & $O(N^2)$ (attn.) & No & No \\
Ensemble~\cite{breiman2001random, Chen_2016} & $O(N \log N)$ & Limited & No \\
MLP & $O(N)$ & No & No \\
\textbf{CoMeT-Net (Ours)} & \textbf{$O(N \cdot C)$} & \textbf{Yes} & \textbf{Yes} \\ \bottomrule
\end{tabular}
\vspace{-0.2cm}
\end{table}


\greenub{\noindent\textbf{CoMeT-Net Novelty:} Unlike reconstruction-based pipeline requiring compression, matching, and reconstruction and transformer-based pipeline (both MemAE and transformer-based methods requires $O(N^2)$ complexity), CoMeT-Net employs a discriminative memory bank with direct template voting, achieving linear complexity $O(N \cdot C)$ where $C$ is the number of classes. Our ResNet backbone extracts $N$ distinct feature patterns (versus single-pattern Multi-Layer-Perceptrons~(MLPs)), enabling fine-grained consensus voting across multiple traffic characteristics. The contrastive alignment loss jointly optimizes memory templates and feature representations, creating interpretable clusters where normal and attack traffic naturally separate. Native E2 integration enables automated PRB throttling and RRC release unavailable in existing methods (Table~\ref{tab:comparison}).}

\section{Proposed Work}\label{sec:proposed_work}
\greenub{To address and meet the aforementioned challenges and requirements,  we propose CoMeT-Net. CoMeT-Net is a discriminative anomaly detection framework for O-RAN (Fig.~\ref{fig:enter-label}), employing template matching with $O(N \cdot C)$ complexity where $N$ feature vectors vote for $C$ classes via cosine similarity. The architecture integrates three innovations: \emph{(1)~structured memory bank} $\mathbf{M} \in \mathbb{R}^{N\times C\times D}$ storing feature-specific templates $\mathbf{M}_{n,c}$ for granular pattern matching, where $D$ denotes the dimensionality of each feature and template vector; \emph{(2)~adaptive gating} outputting weights $g_n$ that downweight ambiguous matches, acting as a learned noise filter; \emph{(3)~contrastive alignment} with margin $m$ structuring the latent space for interpretable boundaries. CoMeT-Edge servers and CoMeT-xApp enable automated PRB throttling and RRC release.}
\noindent\textbf\noindent\emph{\textbf{Memory Bank Formulation:}} \greenub{Network anomalies often manifest in isolated traffic characteristics rather than holistic patterns, a DDoS attack may exhibit abnormal flow rates while maintaining normal packet sizes, or vice versa. Traditional end-to-end classifiers struggle to capture such localized anomalies as they compress all features into a single decision boundary. We address this by decomposing anomaly detection into feature-specific pattern matching, where each traffic characteristic is independently compared against learned templates.}

\greenub{CoMeT-Net's memory bank $\mathbf{M} \in \mathbb{R}^{N\times C\times D}$ stores feature-specific templates that enable this granular detection. Each template $\mathbf{M}_{n,c} \in \mathbb{R}^D$ corresponds to the $n$-th traffic characteristic for class $c$, enabling localized anomaly detection. Conceptually, $\mathbf{M}_{1,\text{DDoS}}$ could capture high-flow-rate signatures while $\mathbf{M}_{2,\text{DDoS}}$ captures packet size distributions, allowing the system to identify DDoS attacks even when only specific traffic aspects are anomalous.}

\greenub{To extract these diverse traffic characteristics, we employ a ResNet-18 backbone that produces $N$ distinct feature vectors $\{\mathbf{E}_n\}_{n=1}^N$, where each $\mathbf{E}_n \in \mathbb{R}^D$ captures different aspects of the input traffic. Unlike MLPs that compress inputs into a single holistic representation, ResNet's convolutional layers with spatial pooling generate $N$ spatially distributed features from different receptive fields. Each feature specializes in specific traffic patterns, enabling the detection of anomalies that appear in isolated characteristics such as abnormal packet timing despite normal flow rates.}


\greenub{Each feature $\mathbf{E}_n$ votes for classes via cosine similarity to its corresponding templates, forming a \textit{voting process} where $N$ diverse perspectives contribute to detection. The score matrix $\mathbf{S} \in \mathbb{R}^{N\times C}$ quantifies these votes using $D$-dimensional vectors:}
$$ S_{n,c} = \frac{\sum_{i=1}^{D} E_{n,i} \times M_{n,c,i}}{\sqrt{\sum_{i=1}^{D} E_{n,i}^2} \times \sqrt{\sum_{i=1}^{D} M_{n,c,i}^2}}\greenub{,} $$
%
%
\greenub{where $S_{n,c} \in [-1, 1]$ measures alignment between feature $n$ and class $c$ template. High $S_{n,c}$ indicates strong pattern match. Direct cosine similarity eliminates additional classifier networks, reducing computational overhead to $O(N \cdot C)$ and enabling sub-millisecond inference critical for O-RAN. Moreover, each $S_{n,c}$ reveals which traffic characteristic voted for which class, providing interpretability for security analysis.}

\noindent\textbf\noindent\emph{\textbf{Gating Mechanism:}} \greenub{While the voting process produces $N$ similarity scores $S_{n,c}$ for each class, these votes vary in reliability. Not all features are equally informative for anomaly detection. During a DDoS attack, flow rate anomalies provide strong discriminative signals while packet timing may remain ambiguous. We introduce an adaptive gating mechanism that acts as a learned noise filter, dynamically weighting each feature's contribution based on its discriminative value.}

\greenub{For each feature vector, the gating network $\phi_{\text{Gate}}$ analyzes its similarity score distribution across classes $S_{n,:} \in \mathbb{R}^C$ and outputs a scalar weight $g_n = \phi_{\text{Gate}}(S_{n,:})$. When $S_{n,:}$ shows similar values across classes, the feature is ambiguous and $\phi_{\text{Gate}}$ assigns low weight. Conversely, when $S_{n,:}$ exhibits large gaps between classes, the feature is discriminative and receives high weight. The weighted scores are:}
$$ S^{\text{Gated}}_{n,c} = S_{n,c} \times g_{n}\greenub{.} $$
\greenub{The gating network $\phi_{\text{Gate}}$ is implemented as a lightweight MLP that learns to identify reliable votes during training. This allows features to modulate their influence or effectively abstain when uncertain.}

\greenub{The final class scores aggregate weighted votes: $L_{c} = \sum_{n=1}^{N} S^{\text{Gated}}_{n,c}$. By downweighting ambiguous matches and emphasizing discriminative ones, the gating mechanism reduces false alarms while maintaining high detection rates. This noise filtering is critical for O-RAN resource control, where misclassifications trigger unnecessary PRB throttling that degrades QoS for legitimate users.}

\noindent\textbf\noindent\emph{\textbf{Optimization Objective:}} \greenub{To jointly train the ResNet backbone, memory templates $\mathbf{M}$, and gating network $\phi_{\text{Gate}}$, we employ a unified loss combining classification and contrastive alignment:}
$$
\greenub{L_{\text{total}} = L_{\text{cls}} + L_{\text{contrastive}}.}
$$
\greenub{The classification loss $L_{\text{cls}}$ optimizes the aggregated scores $L_c$ from the gating mechanism for correct predictions. Given batch size $B$, predicted probabilities $p_{b,c}$, and ground-truth labels $y_{b,c} \in \{0,1\}$:}
$$
\greenub{L_{\text{cls}} = -\frac{1}{B} \sum_{b=1}^{B} \sum_{c=1}^{C} y_{b,c} \log(p_{b,c}).}
$$
\greenub{This cross-entropy loss ensures discriminative detection at the final output level. However, optimizing only the final output does not guarantee that the underlying features are properly structured for template-based matching, potentially creating misalignment between feature representations and the memory bank's similarity-based decision mechanism.}

\greenub{The contrastive alignment loss addresses this by structuring the latent space through margin-based separation between correct and incorrect template matches. For each feature vector, it maximizes similarity to the correct class template while minimizing similarity to negative class templates. With $S_{b,n,c}$ denoting the similarity between feature $n$ of sample $b$ and template for class $c$, $c_{\text{true}}(b)$ the ground-truth class, and $c_{\text{neg}}$ negative classes:}
$$
\greenub{L_{\text{contrastive}} = \frac{1}{B} \sum_{b=1}^{B} \sum_{n=1}^{N} \sum_{c_{\text{neg}}} \max\left(0, m + S_{b,n,c_{\text{neg}}} - S_{b,n,c_{\text{true}}(b)}\right).}
$$
\greenub{The margin parameter $m$ enforces minimum separation between positive and negative matches. The hinge loss $\max(0, \cdot)$ activates only when separation falls below $m$, ensuring well-separated class clusters in the feature space.}

\greenub{Operating at the feature level rather than only at the final output provides critical benefits. The contrastive loss guides the ResNet backbone to extract features naturally aligned with template-based classification, eliminates the disconnect between feature learning and detection objectives, and creates interpretable feature spaces where template similarity directly indicates class membership. This joint optimization of all components enables robust detection of subtle anomalies that might evade classifiers trained solely on cross-entropy loss.}

\noindent\textbf\noindent\emph{\textbf{Threat Mitigation and Dynamic Resource Management:}} \greenub{After training CoMeT-Net with the unified optimization objective, we deploy it within O-RAN infrastructure to enable real-time threat mitigation through coordinated operation of multiple network components.}

\greenub{CoMeT-Edge servers are co-located with UPF at the network edge, positioning them at the critical juncture where user traffic enters the core network. Each CoMeT-Edge server passively monitors IP flows traversing the UPF, extracting traffic features including protocol type (TCP/UDP/ICMP), inferred service from port mappings (HTTP, DNS, FTP), TCP control flags (SYN, ACK, FIN, RST), payload sizes, and inter-packet timing characteristics. After preprocessing through categorical encoding and numeric scaling, the server feeds these features to CoMeT-Net for per-packet classification through the ResNet → Memory → Gating pipeline. To stabilize detection and avoid transient false alarms, the server aggregates classifications over sliding windows of $P$ packets and a stride of $Z$, computing anomaly ratio $r_i = A_i/(N_i + A_i) \in [0,1]$ for each user $i$, where $N_i$ and $A_i$ denote normal and anomalous packet counts respectively.}

\greenub{The Near-RT RIC, a platform that runs xApps for RAN optimization and control with control loop periods of 10ms to 1s, hosts the CoMeT-xApp as the centralized decision-making component for resource control. CoMeT-Edge servers continuously transmit anomaly ratios $r_i$ tagged with slice identifiers via Single Network Slice Selection Assistance Information (S-NSSAI), a standardized tuple that uniquely identifies network slices and their service characteristics (e.g., enhanced mobile broadband, ultra-reliable low-latency), to the xApp through persistent secure TCP connections. Network slicing enables multiple virtualized and isolated logical networks to run on shared physical infrastructure, each tailored to specific service requirements with dedicated resource allocations, QoS guarantees, and security policies. The xApp aggregates reports from all edge servers across network slices, providing a network-wide view of threat distribution. Based on these real-time anomaly ratios, the xApp computes dynamic PRB quotas for each user, where PRBs are the fundamental time-frequency resource units in 5G NR that determine available bandwidth. For partially anomalous users ($0 < r_i < 1$), provisional quotas are calculated as $q_i^{\text{raw}} = \lfloor 100(1-r_i) \rfloor$, inversely proportional to anomaly severity, allowing graduated throttling while maintaining partial service. For fully malicious users ($r_i = 1$), the xApp sets quotas to 0\% and generates RRC connection release commands, which terminate the radio connection and force the UE to detach from the network, preventing further resource consumption. When total demand exceeds capacity ($\sum_i q_i^{\text{raw}} > 100$), the xApp applies proportional scaling $\alpha = 100 / \sum_i q_i^{\text{raw}}$ to ensure fair resource distribution without oversubscription.}

\greenub{The xApp enforces resource control decisions through the O-RAN E2 interface, which connects the Near-RT RIC to gNBs which are the 5G base stations responsible for radio resource management, user scheduling, and radio protocol termination. Using E2 Service Model—RAN Control (E2SM-RC), a standardized service model that defines control procedures for modifying RAN parameters (e.g., admission control, resource allocation, mobility management), the xApp constructs control messages that bind computed PRB quotas to specific network slices identified by S-NSSAI. The gNB receives these messages, updates its scheduler configuration to enforce per-slice PRB limits, and executes RRC connection releases for fully malicious users through its RRC protocol stack. As anomalous users are throttled or disconnected, the gNB scheduler automatically reallocates freed PRBs to legitimate users according to standard scheduling policies, maintaining QoS for benign traffic.}

\greenub{The system operates as a continuous control loop spanning edge detection, centralized decision-making, and distributed enforcement. CoMeT-Edge servers perform local anomaly detection with sub-millisecond latency, the Near-RT RIC xApp computes network-wide resource allocations each control epoch, and gNBs enforce these allocations through standardized interfaces. This architecture decouples detection (edge servers), control (xApp), and enforcement (gNBs), enabling scalable deployment where multiple edge servers feed a single xApp coordinating multiple gNBs. The continuous adaptation to evolving threat patterns ensures resilient network operation while isolating malicious traffic from legitimate users.}

\begin{table}[!t]
\centering
\caption{\label{table_specs} Specifications of the tested hardware platforms.}
\renewcommand{\arraystretch}{1.1}
\setlength{\tabcolsep}{2pt}
\scriptsize
\begin{tabular}{c|c|c|c|c|c}
\hline
\textbf{Platform} & \textbf{Picture} & \textbf{CPU} & \textbf{Clock} & \textbf{GPU} & \textbf{Mem.} \\ 
\hline
Server &
\includegraphics[width=0.07\columnwidth]{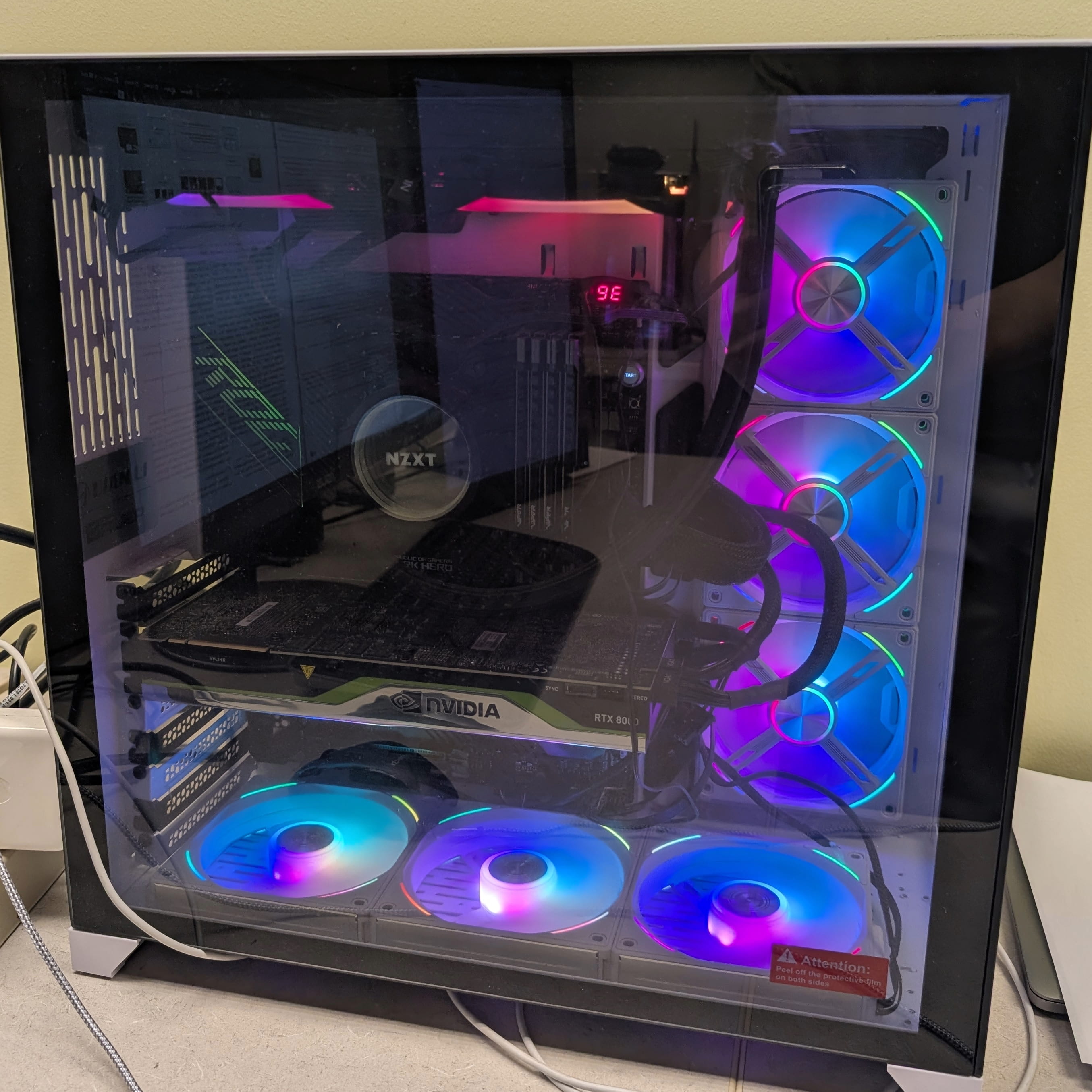} &
AMD Ryzen 5950x & 3.4 GHz & RTX 8000 & 128GB \\
\hline
Laptop &
\includegraphics[width=0.07\columnwidth]{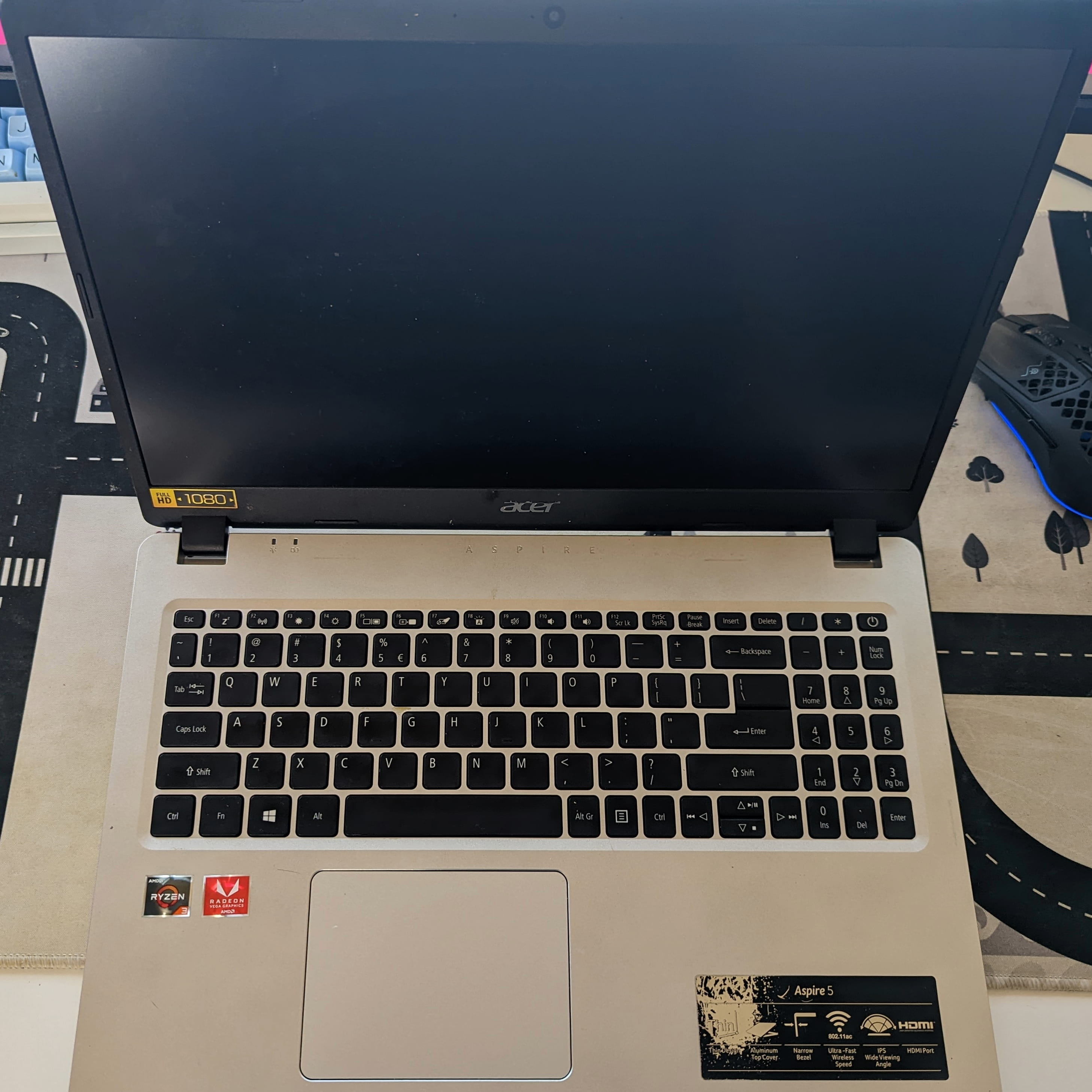} &
AMD Ryzen 3200U & 2.6 GHz & N/A & 4GB \\
\hline
Raspberry Pi 4 &
\includegraphics[width=0.07\columnwidth]{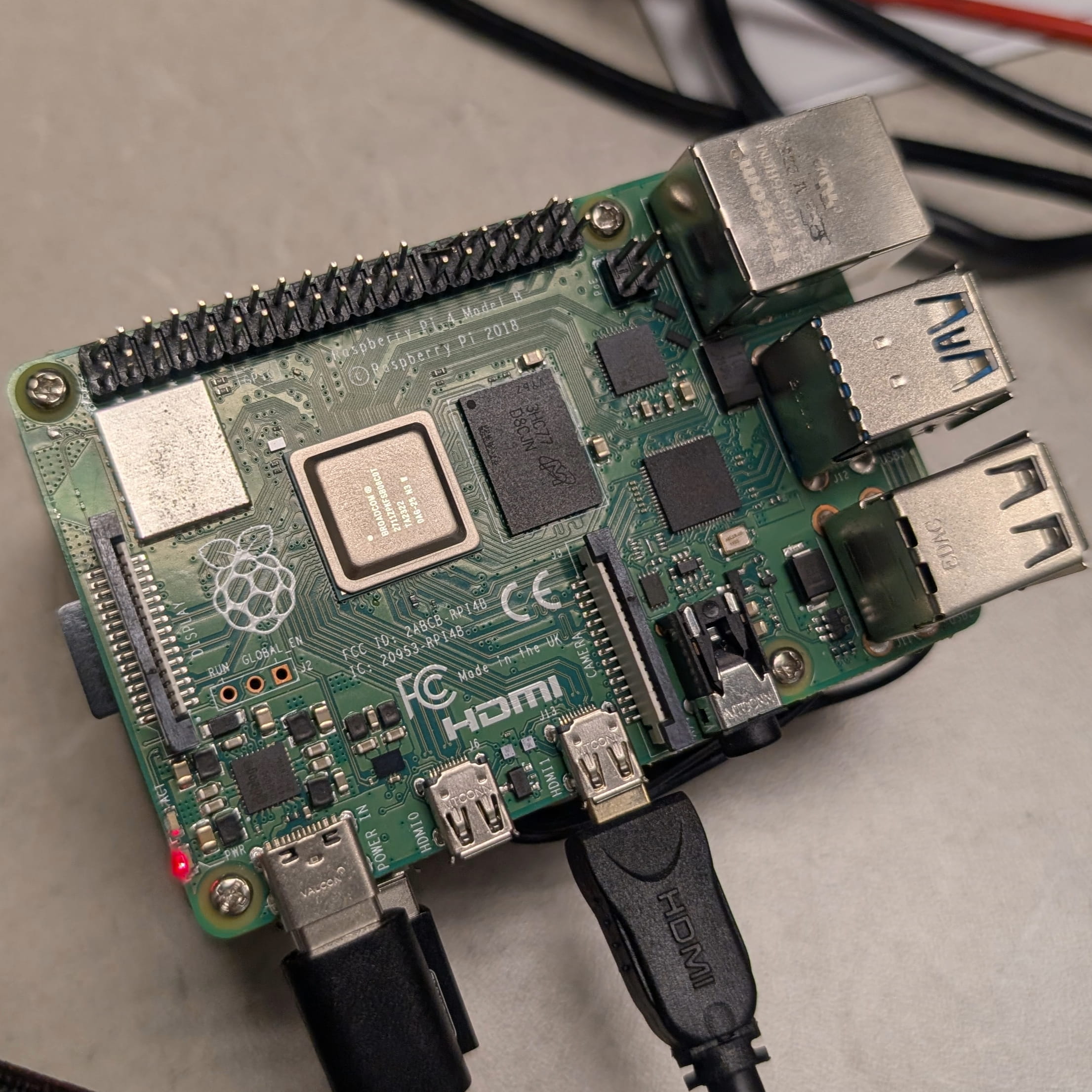} &
Broadcom 2711 & 1.8 GHz & N/A & 8GB \\
\hline
Jetson AGX Xavier &
\includegraphics[width=0.07\columnwidth]{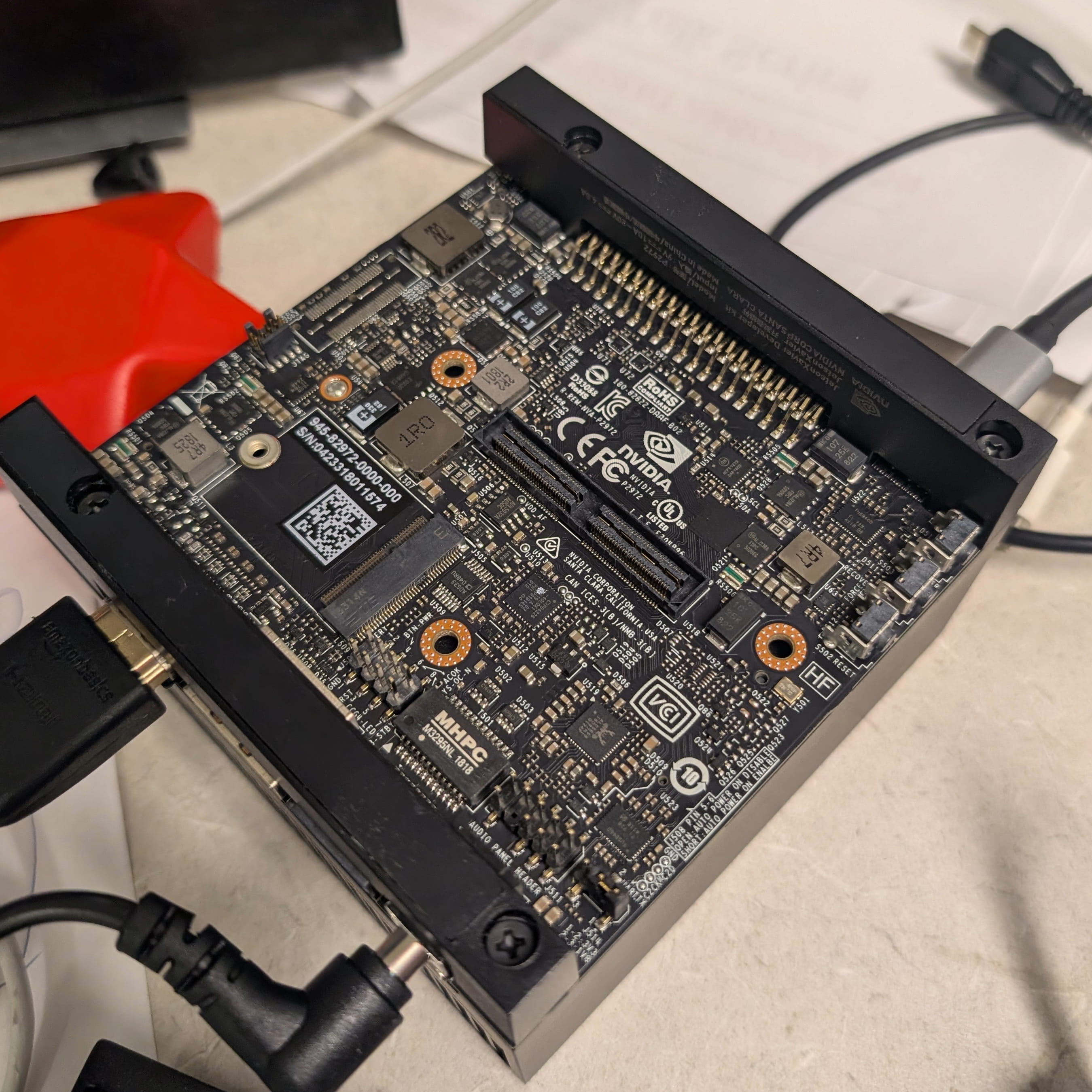} &
NVIDIA Carmel & 1.9 GHz & NVIDIA Volta & 16GB \\
\hline
\end{tabular}
\vspace{-0.6cm}
\end{table}

\begin{table*}[ht!]
\caption{Results on NF-UNSW-NB15-v2~\cite{sarhan2023nf}. CoMeT-Net achieves the best F1 score (99.35\%) with 10× lower false alarm rate (0.0003) compared to baselines, demonstrating effectiveness for network anomaly detection.}
\label{tab:NB15_V2}
\centering
\begin{tabular}{@{}lllllllllll@{}}
\toprule
\textbf{Metric}           & \textbf{LR} & \textbf{NB} & \textbf{SVM} & \textbf{MLP} & \textbf{RF} & \textbf{LGB} & \textbf{XGB} & \textbf{CatB} & \textbf{FTT} & \textbf{CoMeT-Net} \\ \midrule
\textbf{Precision}        & 0.8394      & 0.7583      & 0.8602       & 0.8835       & 0.9134      & 0.9035       & 0.9064       & 0.9026        & 0.7439                 & 0.9943             \\
\textbf{Recall}           & 0.9935      & 0.9999      & 0.9972       & 0.9920       & 0.9893      & 0.9912       & 0.9900       & 0.9885        & 1.0000                 & 0.9933             \\
\textbf{F1 Score}         & 0.9100      & 0.8625      & 0.9236       & 0.9346       & 0.9499      & 0.9453       & 0.9464       & 0.9436        & 0.8531                 & 0.9935             \\
\textbf{False Alarm Rate} & 0.0079      & 0.0132      & 0.0067       & 0.0054       & 0.0039      & 0.0044       & 0.0042       & 0.0044        & 0.0143                 & 0.0003             \\ \bottomrule
\end{tabular}
\end{table*}

\begin{table*}[]
\caption{Results obtained on the Backdoor dataset. With near-perfect balance between precision (98.67\%) and recall (95.14\%), CoMeT-Net delivers excellent performance (96.88\% F1 score) on cybersecurity backdoor detection, validating our gating mechanism's effectiveness at identifying subtle malicious patterns.}
\label{tab:backdoor}
\centering
\begin{tabular}{@{}lllllllllll@{}}
\toprule
\textbf{Metric}           & \textbf{LR} & \textbf{NB} & \textbf{SVM} & \textbf{MLP} & \textbf{RF} & \textbf{LGB} & \textbf{XGB} & \textbf{CatB} & \textbf{FTT} & \textbf{CoMeT-Net} \\ \midrule
\textbf{Precision}        & 0.9954      & 1.0000      & 0.9867       & 0.9848       & 0.9932      & 0.9830       & 0.9882       & 0.9907        & 0.9779                 & 0.9867             \\
\textbf{Recall}           & 0.8389      & 0.7609      & 0.8517       & 0.9130       & 0.9399      & 0.9616       & 0.9668       & 0.9488        & 0.9054                 & 0.9514             \\
\textbf{F1 Score}         & 0.9105      & 0.8642      & 0.9142       & 0.9476       & 0.9658      & 0.9722       & 0.9774       & 0.9693        & 0.9402                 & 0.9688             \\
\textbf{False Alarm Rate} & 0.0001      & 0.0000      & 0.0003       & 0.0004       & 0.0002      & 0.0004       & 0.0003       & 0.0002        & 0.0005                 & 0.0003             \\ \bottomrule
\end{tabular}
\vspace{-0.2cm}
\end{table*}
\section{Performance Evaluation }\label{sec:performance_eval}
\greenub{We evaluate CoMeT-Net through multi-dataset experiments and O-RAN testbed deployment, addressing three validation objectives: (1) assessing performance on network datasets (NF-UNSW-NB15-v2~\cite{sarhan2023nf}, Backdoor~\cite{han2022adbench}) to validate O-RAN requirements of high-accuracy detection with low false alarms for resource control; (2) demonstrating generalization and deployment on resource-constrained edge hardware, addressing SECON's focus on machine learning for networking; (3) evaluating robustness on a non-network dataset (Campaign~\cite{han2022adbench}) to verify that Memory + Gating generalizes beyond networking, showing potentials in broader applications. Results show 99.35\% F1 score on NB15-v2 with 10$\times$ lower false alarm rates (0.0003 vs. 0.0039), and 0.3-3ms inference across hardware tiers. We also deploy the complete system on an O-RAN testbed to validate real-world applicability.}

\subsection{Experiment Setup and Datasets}
\greenub{CoMeT-Net is implemented with PyTorch~\cite{paszke2019pytorchimperativestylehighperformance} using Adam optimizer~\cite{kingma2014adam} (learning rate $0.001$, batch size $32$), with a ResNet-like architecture adapted for tabular data (1D) comprising $6$ blocks, each containing two convolutional layers and two batch normalization layers. We compare against baseline and state-of-the-art methods from ADBench~\cite{han2022adbench}: traditional machine learning approaches including Logistic Regression~(LR)~\cite{nick2007logistic}, Naive Bayes (NB)~\cite{zhang2004optimality}, and Support Vector Machine (SVM)~\cite{scholkopf2001estimating}; neural networks including Multi-Layer Perceptron (MLP)~\cite{hinton1990connectionist}; ensemble methods including Random Forest (RF)~\cite{breiman2001random}, LightGBM (LGB)~\cite{ke2017lightgbm}, XGBoost (XGB)~\cite{Chen_2016}, and CatBoost (CatB)~\cite{prokhorenkova2018catboost}; and the transformer-based FTTransformer (FTT)~\cite{gorishniy2021revisiting}.}

\greenub{Our multi-dataset evaluation serves three distinct validation objectives aligned with O-RAN deployment requirements. \textbf{NF-UNSW-NB15-v2}~\cite{sarhan2023nf} ($2.39$ million samples, $43$ features, $3.98\%$ anomaly rate) validates core O-RAN requirements: high-accuracy detection with minimal false alarms for resource control, where features (protocol types, payload sizes, inter-packet timing) match those extractable from O-RAN user-plane traffic. Although not collected from O-RAN infrastructure, its attack signatures and categories also presents in the threats O-RAN networks face. \textbf{Backdoor} ($95,329$ samples, $196$ features, $2.44\%$ anomaly rate) demonstrates generalization across diverse network intrusion datasets and validates deployability on our O-RAN testbed, where the model detects attacks attempting to bypass security mechanisms. \textbf{Campaign} ($41,188$ samples, $62$ features, $11.27\%$ anomaly rate from successful bank marketing outcomes) tests robustness of the Memory + Gating mechanism on highly imbalanced non-network data, verifying that our template-based approach generalizes beyond networking domains to support broader pervasive computing applications.} 

\begin{figure*}[]
    \centering
    \begin{subfigure}[b]{0.245\textwidth}
        \centering
        \includegraphics[width=\textwidth]{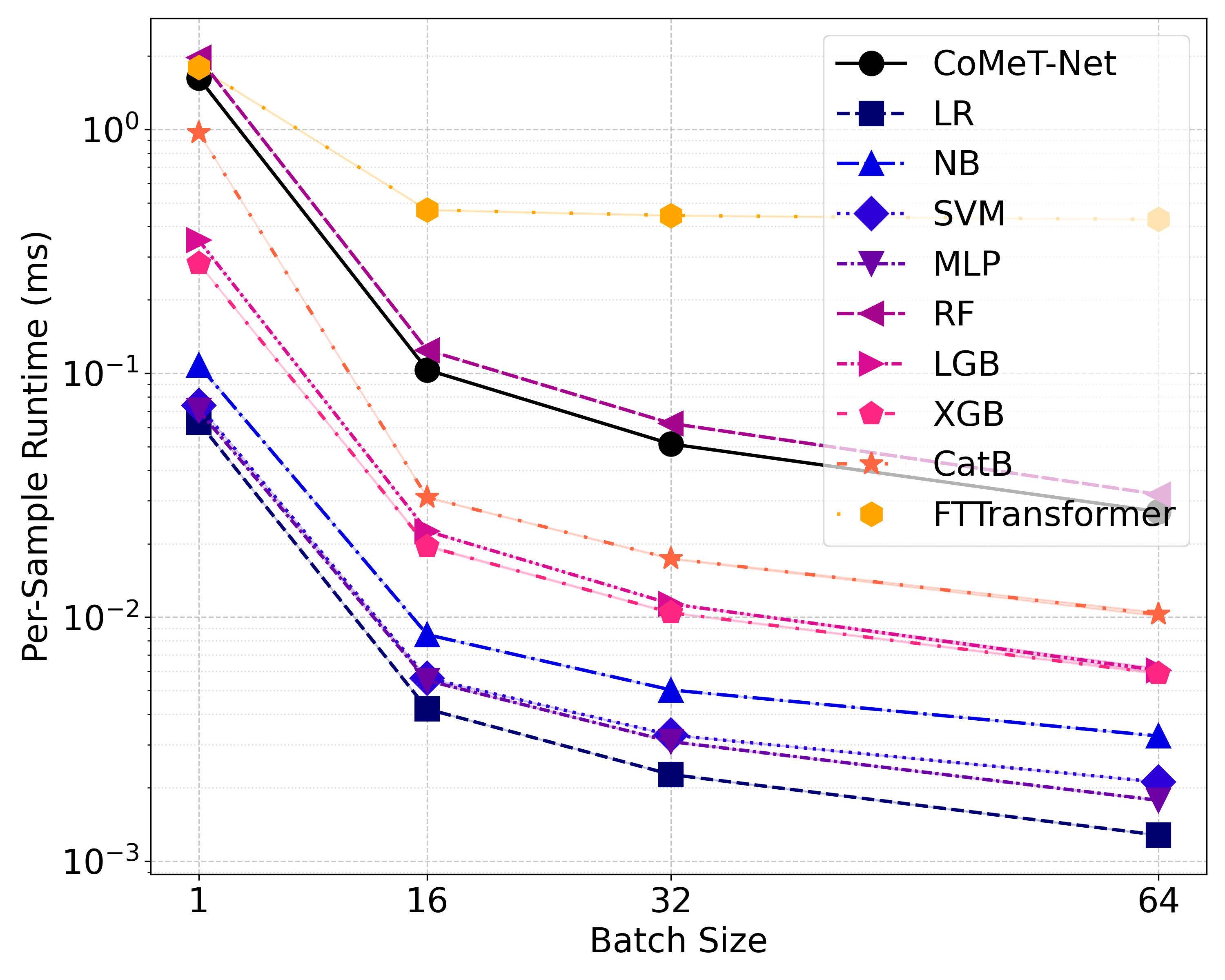}
        \caption{}
        \label{fig:runtime_workstation}
    \end{subfigure}
    \hfill
    \begin{subfigure}[b]{0.245\textwidth}
        \centering
        \includegraphics[width=\textwidth]{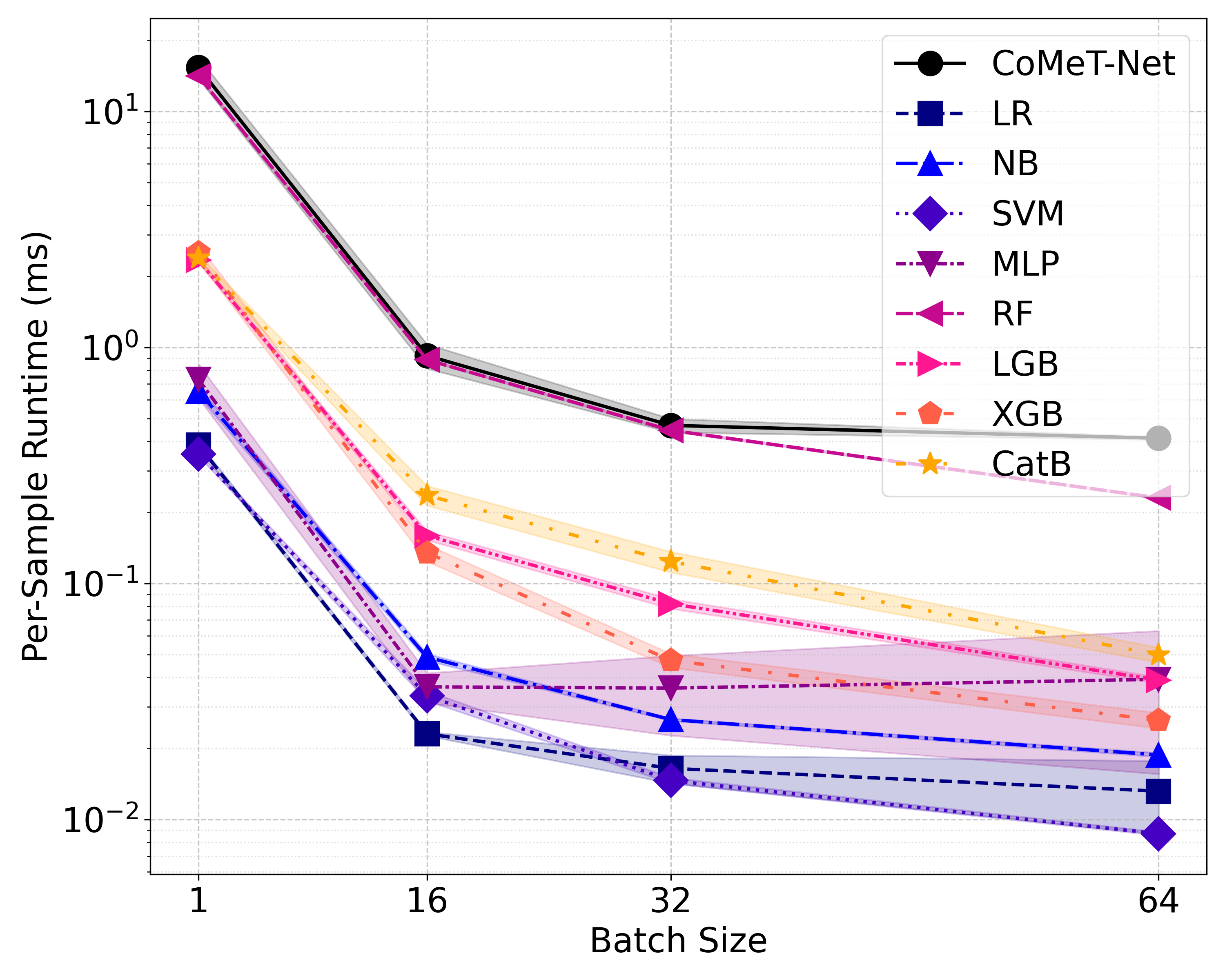}
        \caption{}
        \label{fig:runtime_xavier}
    \end{subfigure}
    \hfill
    \begin{subfigure}[b]{0.245\textwidth}
        \centering
        \includegraphics[width=\textwidth]{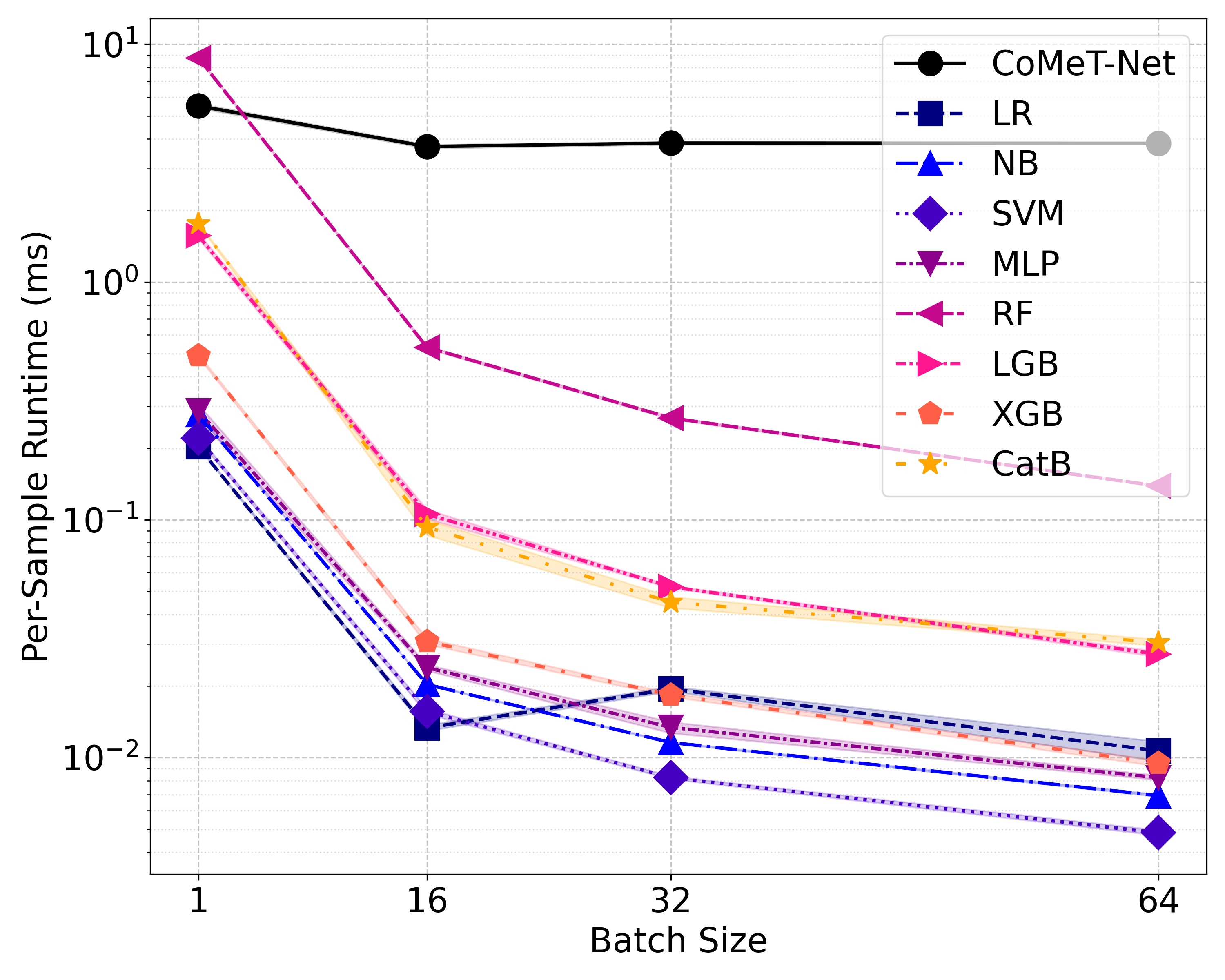}
        \caption{}
        \label{fig:runtime_laptop}
    \end{subfigure}
    \hfill
    \begin{subfigure}[b]{0.245\textwidth}
        \centering
        \includegraphics[width=\textwidth]{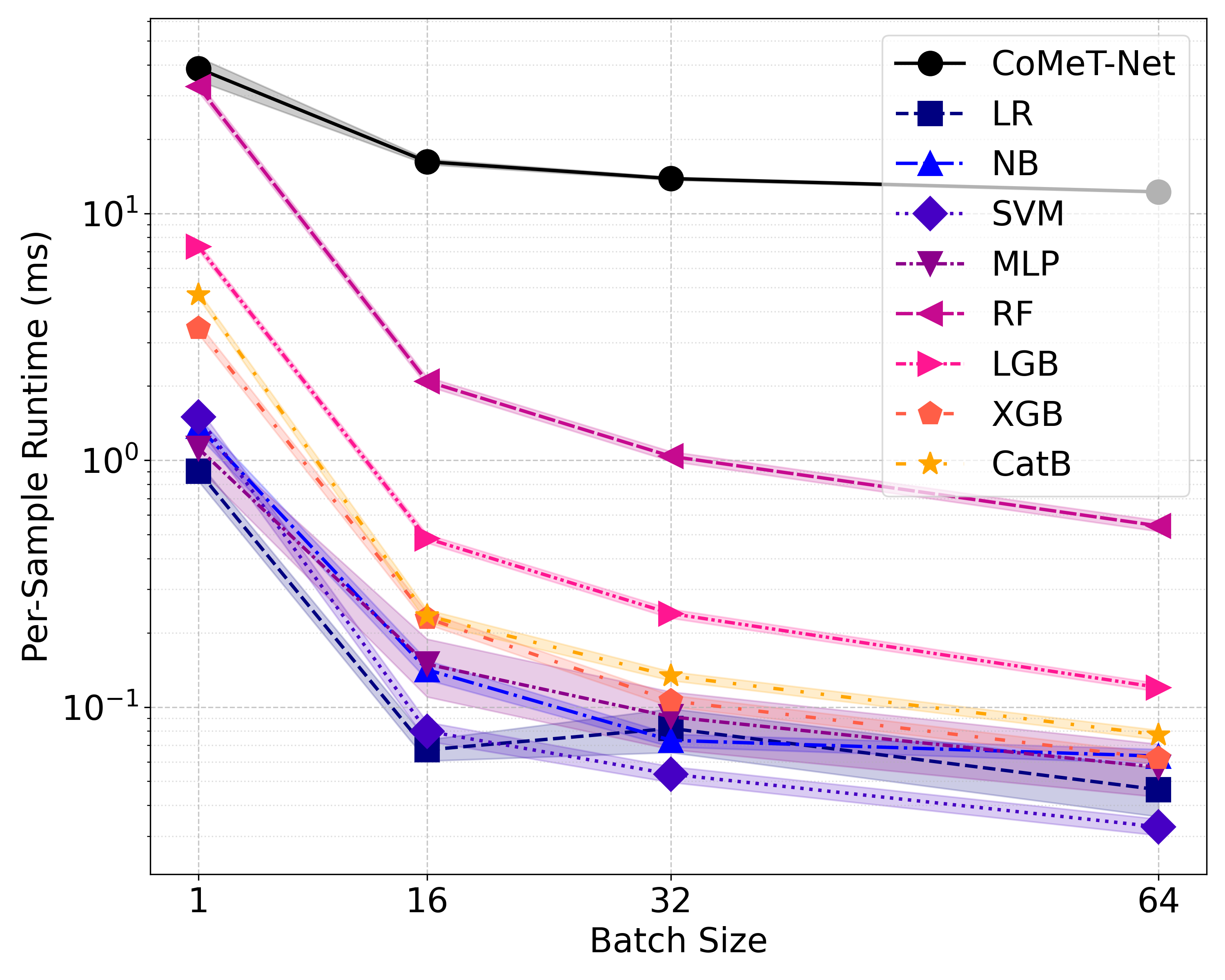}
        \caption{}
        \label{fig:runtime_pi}
    \end{subfigure}
    \caption{Inference time comparison across hardware platforms. Our approach maintains real-time performance across (a) high-performance servers (0.3ms per sample), (b) edge devices (0.5ms), (c) standard laptops (1ms), and (d) resource-constrained Raspberry Pi 4 (under 3ms). CoMeT-Net scales efficiently with batch size while maintaining superior detection accuracy compared to traditional methods, demonstrating practical deployability across the entire computing spectrum.}
    \label{fig:runtime_analysis}
\end{figure*}



\greenub{We evaluated performance using four key metrics. Precision measures the proportion of correctly identified anomalies among all detected instances. Recall quantifies the proportion of actual anomalies successfully identified. The F1 score balances precision and recall. The false alarm rate measures normal instances incorrectly classified as anomalies, which is critical for operational deployment. To evaluate practical implementation feasibility, we profiled runtime on four hardware platforms: a high-performance workstation, an edge AI accelerator (Jetson AGX Xavier), a low-power laptop, and an IoT device (Raspberry Pi 4), shown in Table~\ref{table_specs}. We used an input size of 400 and measured per-sample inference time across batch sizes from 1 to 64 to evaluate scalability}


\noindent\textbf{Networking Dataset Analysis:}
\greenub{CoMeT-Net demonstrates exceptional performance on NF-UNSW-NB15-v2 (Table~\ref{tab:NB15_V2}), achieving F1 score $0.9935$, outperforming other state-of-the-art methods. Our approach maintains excellent recall ($0.9933$) while dramatically improving precision ($0.9943$), addressing a key challenge in network intrusion detection. Critically, the false alarm rate is reduced by an order of magnitude ($0.0003$ vs. $0.0039$), essential for O-RAN environments where false alarms trigger unnecessary PRB reallocation.}

\begin{table*}[ht!]
\caption{Results obtained on the Campaign dataset. Our consensus voting architecture outperforms all baseline methods on financial marketing data with the highest F1 score ($60.30\%$) and exceptional recall ($79.03\%$), showcasing our model's ability to adapt to imbalanced, non-networking domains without specialized tuning.}
\label{tab:campaign}
\centering
\begin{tabular}{@{}lllllllllll@{}}
\toprule
\textbf{Metric}           & \textbf{LR} & \textbf{NB} & \textbf{SVM} & \textbf{MLP} & \textbf{RF} & \textbf{LGB} & \textbf{XGB} & \textbf{CatB} & \textbf{FTT} & \textbf{CoMeT-Net} \\ \midrule
\textbf{Precision}        & 0.6450      & 0.3204      & 0.6384       & 0.5240       & 0.6214      & 0.6178       & 0.6021       & 0.6171        & 0.5681                 & 0.4875             \\
\textbf{Recall}           & 0.3955      & 0.6631      & 0.2656       & 0.5166       & 0.4754      & 0.5533       & 0.5419       & 0.5419        & 0.5053                 & 0.7903             \\
\textbf{F1 Score}         & 0.4903      & 0.4320      & 0.3752       & 0.5203       & 0.5387      & 0.5838       & 0.5704       & 0.5771        & 0.5349                 & 0.6030             \\
\textbf{False Alarm Rate} & 0.0270      & 0.1748      & 0.0187       & 0.0583       & 0.0360      & 0.0425       & 0.0445       & 0.0418        & 0.0477                 & 0.1032             \\ \bottomrule
\end{tabular}
\vspace{-0.3in}
\end{table*}

\greenub{Traditional methods struggle with NB15-v2's complexity. LR achieves high recall ($0.9935$) but poor precision ($0.8394$), producing many false positives, while NB shows the highest recall ($0.9999$) but lowest precision ($0.7583$). Despite its advanced architecture, FTT performs poorly (F1: $0.8531$) due to quadratic self-attention complexity on high-volume traffic. CoMeT-Net's success stems from template vectors modeling distinct traffic patterns and adaptive gating that weights feature contributions, avoiding the equal-importance treatment of traditional methods and enabling subtle attack detection without overgeneralization.}

\greenub{On the Backdoor dataset (Table~\ref{tab:backdoor}), CoMeT-Net achieves strong performance (F1: $0.9688$) with balanced precision ($0.9867$) and recall ($0.9514$). Traditional methods struggle with recall (LR: $0.8389$, NB: $0.7609$), missing many backdoor attacks. CoMeT-Net's ability to capture subtle patterns across multiple feature dimensions proves crucial for detecting attacks with minimal differences from legitimate requests.}

\noindent\textbf{Cross-Domain Dataset Analysis:}
\greenub{CoMeT-Net demonstrates strong cross-domain generalization without domain-specific customization. On the Campaign dataset (Table~\ref{tab:campaign}), it achieves the highest F1 score ($0.6030$) with strong recall ($0.7903$), identifying rare successful campaigns that other methods miss. While SVM achieves high precision ($0.6384$), its poor recall ($0.2656$) misses many anomalies. CoMeT-Net succeeds through contrastive alignment that maximizes class separability and joint feature-detection optimization, proving valuable for financial domains where anomalous patterns share characteristics with normal data. This validates our template matching and consensus voting mechanism's adaptability across distributions while maintaining accuracy and low false alarm rates.}

\noindent\textbf{Generalizability Analysis:}
\greenub{CoMeT-Net achieves consistent top-tier performance across datasets, with exceptional results on NF-UNSW-NB15-v2 (F1: $0.9935$) validating its network intrusion detection capabilities. Unlike baselines that excel in either precision or recall, CoMeT-Net maintains favorable balance through unified optimization while achieving low false alarm rates critical for O-RAN deployment. Traditional methods (SVM, RF, FTT) show inconsistent cross-dataset performance, struggling on datasets where they lack task-specific optimization. CoMeT-Net's robustness stems from three architectural innovations: the memory bank adapts to diverse distributions without domain-specific tuning, the gating mechanism performs automated feature selection, and contrastive alignment creates discriminative representations regardless of data characteristics.}

\noindent\textbf{Runtime Analysis:}
Fig.~\ref{fig:runtime_analysis} presents inference times across batch sizes $1-64$ on four hardware platforms representing different deployment scenarios: servers for high-performance computing at base stations, Xavier for distributed O-RAN edge servers, and laptop/Raspberry Pi for resource-constrained devices with different computational and memory limitations.

While simpler models (LR, NB, SVM) achieve lowest absolute inference times, CoMeT-Net demonstrates real-time capabilities across all platforms. On GPU-accelerated systems, CoMeT-Net achieves inference times comparable to ensemble methods while delivering substantially better detection performance. The high-performance workstation achieves $0.3ms$ per sample, enabling thousands of samples per second processing critical for high-throughput O-RAN deployments. The Xavier platform maintains competitive performance with efficient batch scaling, while CPU-only environments achieve under $1ms$ on laptop and approximately 3ms on Raspberry Pi 4, well within real-time constraints for network traffic processing.

FTTransformer demonstrates severe deployment limitations, failing on Xavier, laptop, and Raspberry Pi due to out-of-memory errors, confirming our critique of transformer approaches where quadratic self-attention complexity creates memory and processing bottlenecks. Ensemble methods (RF, LGB, XGB, CatB) exhibit significant computational overhead compared to simpler models while failing to match CoMeT-Net's accuracy on critical network intrusion detection datasets. These benchmarks validate our architectural design, confirming CoMeT-Net achieves practical balance between state-of-the-art detection performance and computational efficiency necessary for real-world deployment.

\begin{figure}[t]
\centering
\includegraphics[width=0.47\textwidth]{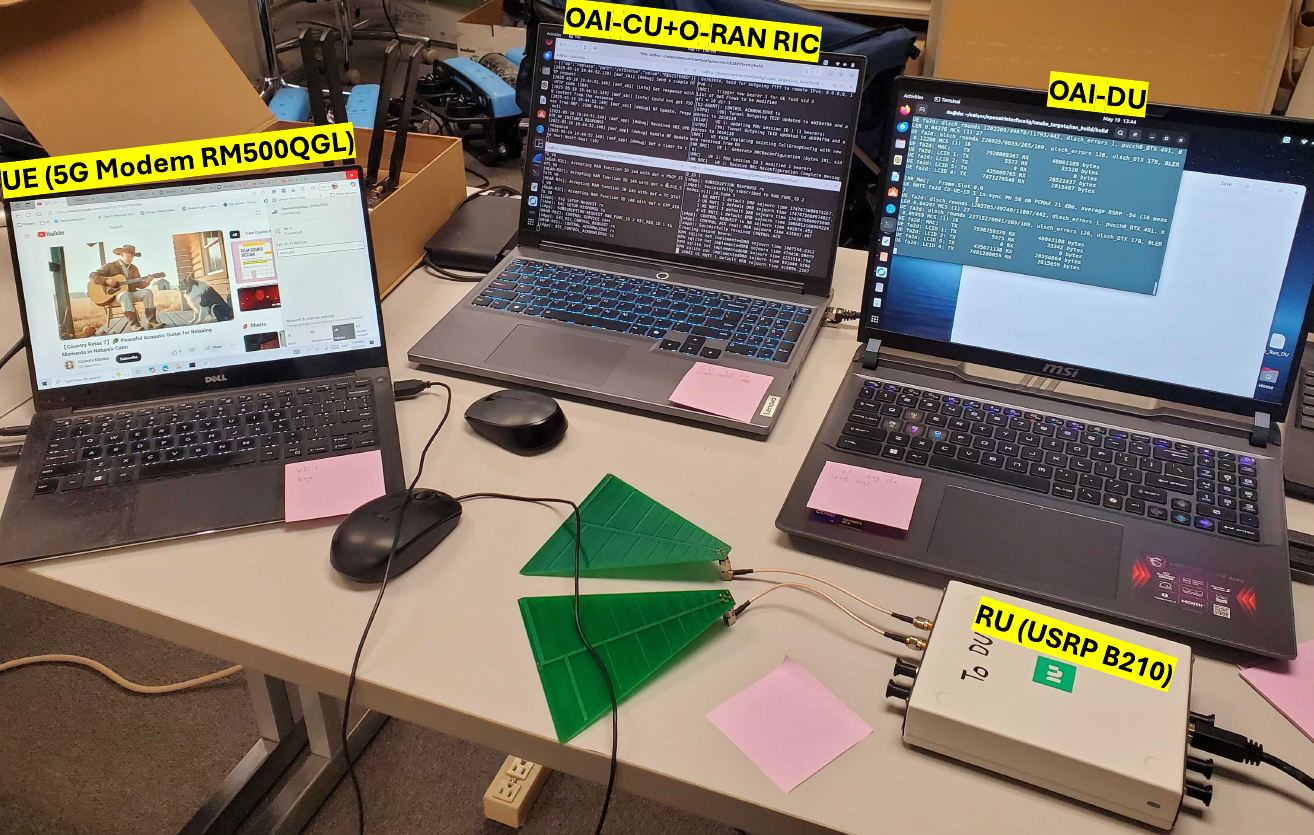}
\caption{O-RAN testbed and hardware setup.}
\label{fig:testbed}
\vspace{-0.15in}
\end{figure}

\subsection{CoMeT-Net O-RAN Testbed Implementation}
We integrate CoMeT-Net into the O-RAN testbed through two primary components: (i) the distributed \emph{CoMeT-Edge servers} positioned at network edges for real-time \emph{($<10$ ms) }traffic analysis and anomaly detection, and (ii) the \emph{CoMeT-xApp}, an extended application deployed in the Near-RT RIC for anomaly-driven dynamic resource reallocation and threat mitigation. Distributed \emph{CoMeT-Edge servers} are co-located with the UPF to enable proximal threat detection and minimize processing latency. Each edge server implements \emph {Lightweight Inference Engine}, Optimized CoMeT-Net models designed for resource-constrained edge environments, maintaining detection accuracy while reducing computational overhead. 

\begin{figure}[t]
\centering
\includegraphics[width=0.92\columnwidth]{\detokenize{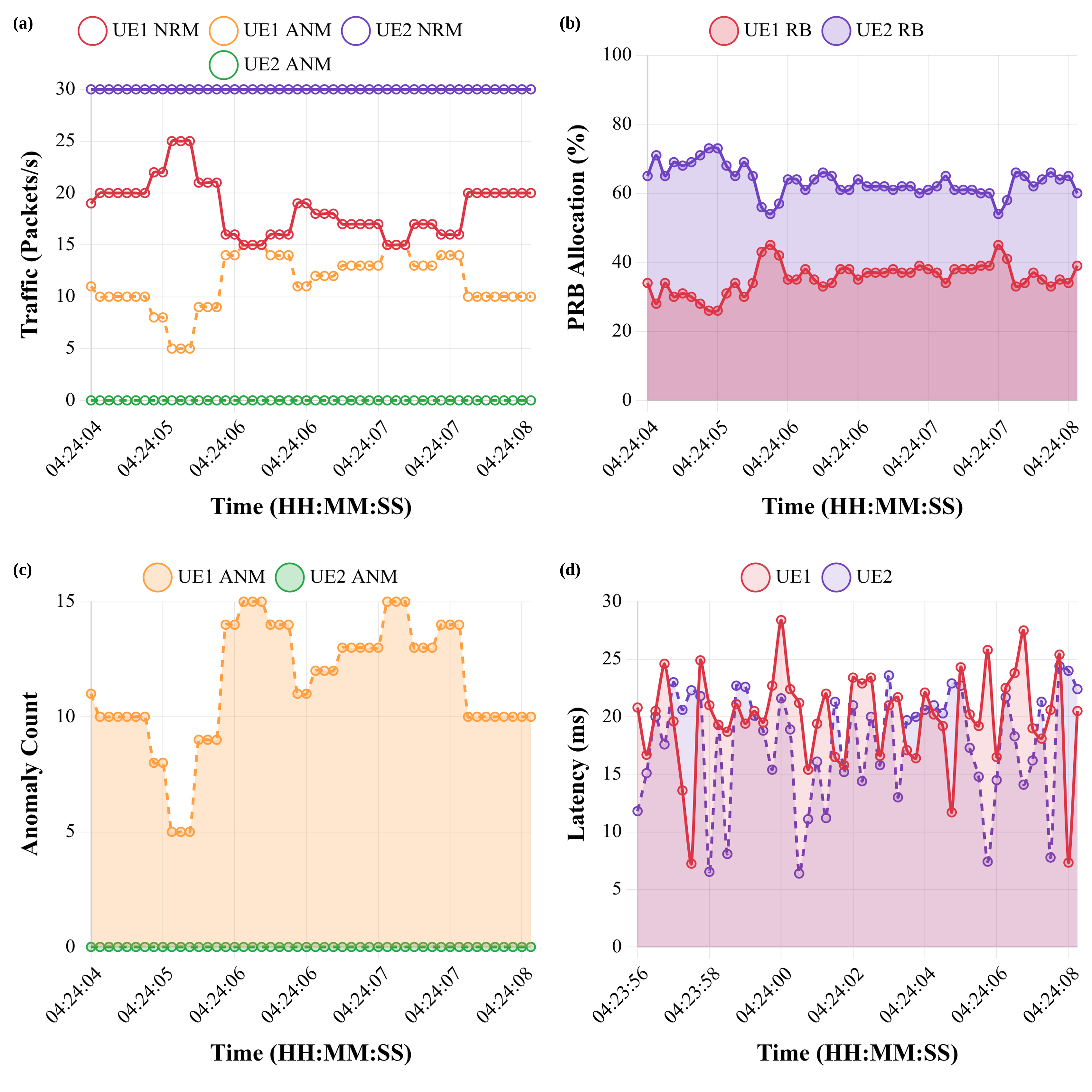}}
\caption{ Real-time Anomaly-driven PRB reallocation: (a) traffic, (b) PRB, (c) anomalies, (d) latency.}
\label{fig:dyn}
\vspace{-0.3in}
\end{figure}

\begin{figure}[t]
\centering
\begin{subfigure}[t]{0.9\linewidth}
\centering
\includegraphics[width=\linewidth]{\detokenize{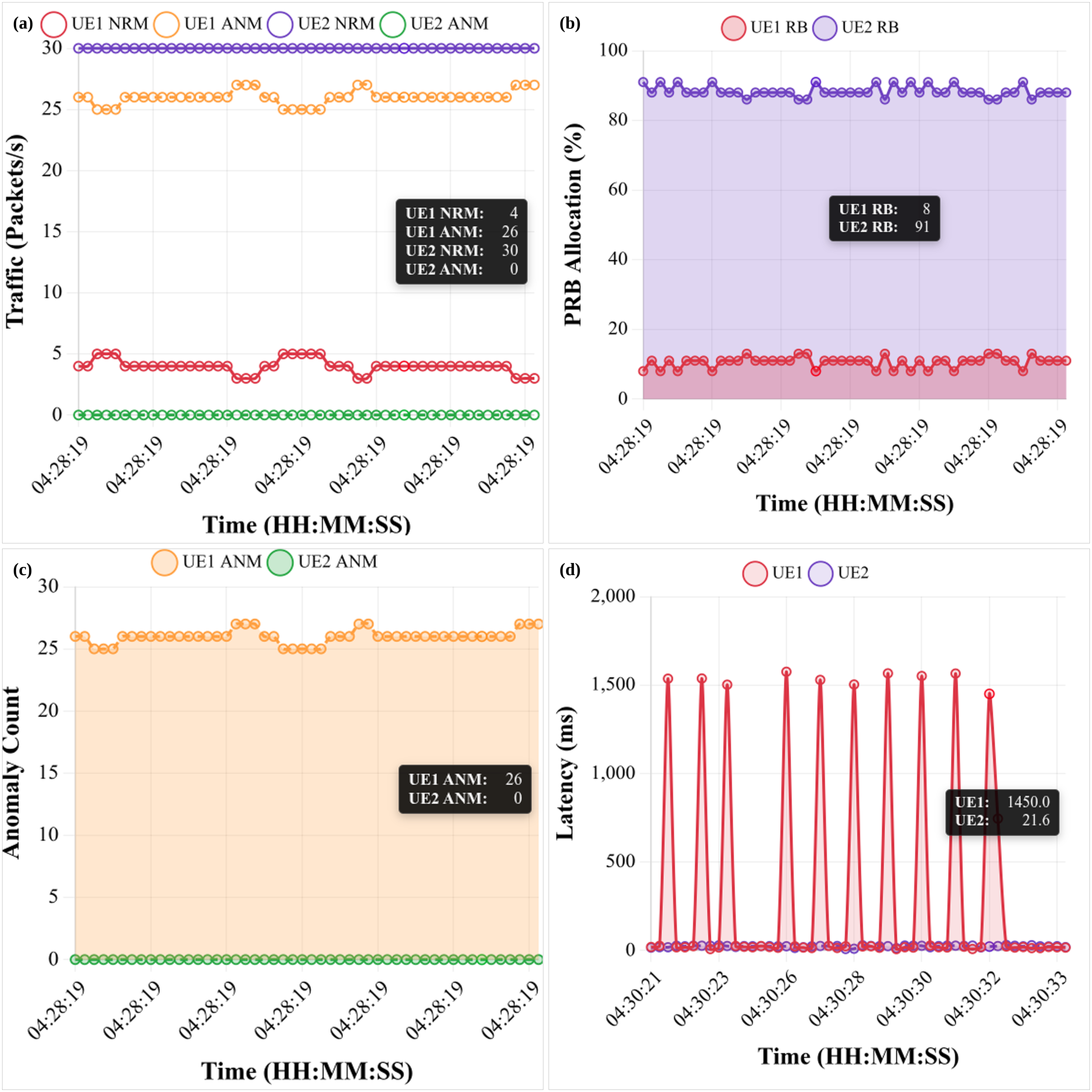}}
\caption{System performance under flood attack.}
\label{fig:att}
\end{subfigure}

\vspace{0.6em}

\begin{subfigure}[t]{0.9\linewidth}
\centering
\includegraphics[width=\linewidth]{\detokenize{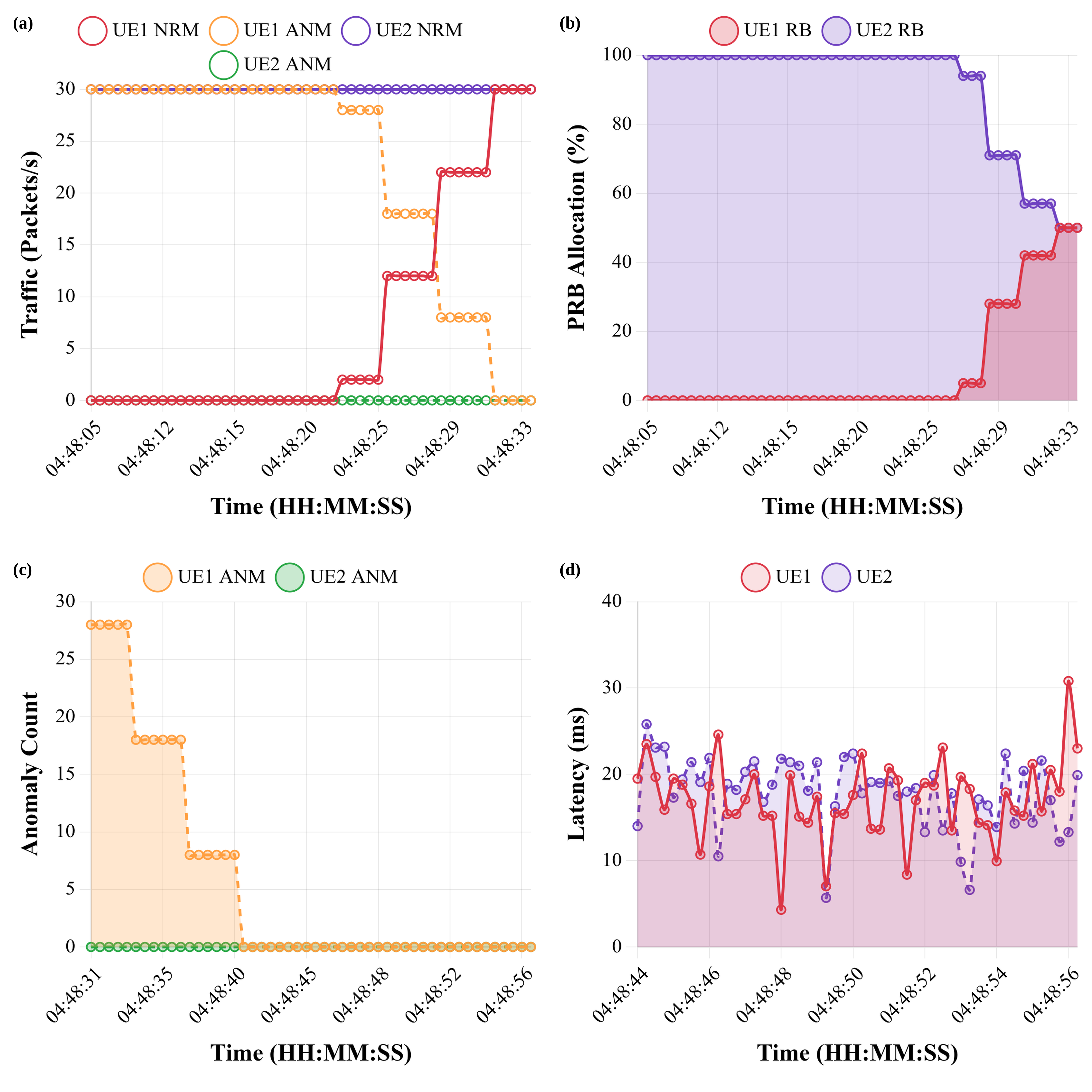}}
\caption{Post-attack recovery.}
\label{fig:nor}
\end{subfigure}

\caption{Attack vs.\ post-attack recovery using CoMeT-Net.}
\label{fig:att-nor}
\vspace{-0.3in}
\end{figure}


\noindent\textbf{Testbed Setup and Integration.}
As shown in Fig.~\ref{fig:testbed}, our O-RAN testbed includes: \emph{i)~gNB}: OAI~\cite{oai} with the Central Unit (CU) and Distributed Unit (DU) deployed on an AMD Ryzen 9 9950X (16c/32t) server with an RTX~5080 (16~GB) and 128~GB RAM; \emph{ii)~RU}: USRP B210 SDRs in Band~78 (3.7~GHz) for over-the-air transmission; \emph{iii)~5G core}: OAI core (AMF, SMF, UPF, and supporting functions); \emph{iv)~O-RAN RIC}: FlexRIC~\cite{FlexRICSoftware} with CoMeT-xApp integration; \emph{v)~UEs}: OAI UEs and commercial modems (RM500Q-GL, EM9191).

\noindent\textbf{Testbed Experimental Results.} 
We evaluated the CoMet-Net framework on a real 5G SDR testbed with two UEs across dynamic, sustained attack, and post-attack recovery scenarios.

\noindent\underline{\emph{\textbf{Dynamic Resource Allocation:}}}  
To evaluate system behavior under dynamic scheduling, both UEs initially generate benign traffic and share PRBs equally ($50\%$ each). Adversarial activity is then introduced by replaying malicious flows from the unseen KDD~\cite{5356528} test split on UE1, while UE2 continues normal operation. At the edge, CoMeT\hyp Net computes the per\hyp UE anomaly ratio $r_i$ and streams it to the Near\hyp RT RIC xApp. The xApp maintains UE\hyp specific data structures (latest $r_i$, state, throttling flags) and computes the instantaneous PRB allocation $P_i(r_i)$ online, enforcing proportional throttling for anomalous UEs and compensatory boosts for benign UEs. Over the observation window $04{:}24{:}04$--$04{:}24{:}08$, UE1 shows intermittent anomalous bursts (ANM $\sim$$10-15~\rm{pkts/s}$) atop normal traffic (NRM $\sim$$15-25~\rm{pkts/s}$), while UE2 sustains a stable $\sim$$30~\rm{pkts/s}$ throughout the experiment (Fig.~\ref{fig:dyn}a). In response to rising $r_1$, the xApp throttles UE1's PRB share to $\sim$$35-40\%$ and boosts UE2's to $\sim$$60-65\%$ (Fig.~\ref{fig:dyn}b), closely mirroring anomaly\hyp count fluctuations (UE1 $\sim$$5-15$ during attack intervals; Fig.~\ref{fig:dyn}c). The impact on QoS is evident in the latency measurements: UE1's latency rises from a $15-20\rm{ms}$ baseline to peaks of $28\rm{ms}$ during throttling (a $\sim$$40\%$ increase), whereas UE2's average latency remains stable at $15-20\rm{ms}$ (Fig.~\ref{fig:dyn}d), demonstrating effective isolation of the malicious user while preserving QoS for legitimate traffic.

\noindent\underline{\emph{\textbf{Sustained Attack:}}} Fig.~\ref{fig:att} shows the system behavior under sustained flood attack conditions. The anomaly detector identifies persistent malicious patterns from UE1, with anomaly counts reaching $26$ (Fig.~\ref{fig:att}(c)). The framework responds with aggressive resource throttling, reducing UE1's PRB allocation from $50\%$ to below $10\%$ within the attack window (Fig.~\ref{fig:att}(b)). This results in severe latency penalties for the attacker, with Round Trip Time (RTT), the time for a packet to travel from source to destination and back, spikes exceeding $1500\rm{ms}$ (Fig.~\ref{fig:att}(d)), effectively containing the attack impact while maintaining UE2's stable operation at $21.6\rm{ms}$ average latency.

\begin{table}[t]
\centering
\caption{Performance with/without CoMeT-Net under attack (UE2 is benign; mean RTT in ms).}
\label{tab:comet-results}
\begin{tabular}{lccc}
\toprule
\textbf{Use CoMeT-Net} & \textbf{UE1 RTT} & \textbf{UE2 RTT} & \textbf{CPU Usage} \\
\midrule
No  & 30.2 & 1446.1 & 48.3\% \\
Yes & 1449.2 &   17.4 &  3.8\% \\
\bottomrule
\end{tabular}
\vspace{-0.3in}
\end{table}


\noindent\underline{\emph{\textbf{Post-Attack Recovery and System Resilience:}}}
We recorded the performance of the UEs in a setting with one attacker (UE1) and one benign user~(UE2). As shown in Table~\ref{tab:comet-results}, in the absence of CoMeT-Net, UE2’s downlink rate degrades sharply after the attack, while UE1 continues to consume significant bandwidth as a result of the attack; RTT averaged around $1450\rm{ms}$ and O-RAN CU's CPU usage averaged $48\%$, indicating failure to isolate the attacker.  The performance did not recover after the attack ended, and UE2 continued to experience consistently high RTT. With CoMeT-Net, isolation and throttling stabilize the system, achieving RTT$\sim$$15$ to $20\rm{ms}$ and CPU$\sim$$3.8\%$. Fig.~\ref{fig:nor} demonstrates the framework's ability to restore baseline performance following anomaly mitigation. The traffic patterns show a marked transition at 04:48:25, where UE1's anomalous behavior terminates, transitioning from elevated rates ($30\rm{pkts/s}$ anomaly traffic) to stable legitimate traffic as shown in (Fig.~\ref{fig:nor}(a)). The anomaly detection mechanism (Fig.~\ref{fig:nor}(c)) registers this behavioral change, with UE1's anomaly count dropping from $27$ to $0$ within seconds of attack termination.

Following the mitigation of malicious behavior, the system triggers resource reconfiguration (Fig.~\ref{fig:nor}(b)). PRB allocation undergoes rapid convergence, transitioning from the attack-induced asymmetric distribution (UE1: $8\%$, UE2: $92\%$) to balanced allocation within a 4-second recovery window (04:48:28-04:48:32). This automated recovery demonstrates the framework's bidirectional adaptability: it constrains resources during attacks and restores fair allocation upon threat mitigation. The latency measurements (Fig.~\ref{fig:nor}(d)) validate this recovery, with both UEs converging to comparable performance levels, averaging $15-20\rm{ms}$, confirming complete restoration of baseline QoS metrics. These results validate the framework's capability to: (i)~detect anomalous behavior in real time with subsecond response times, (ii)~dynamically adjust resource allocation proportional to threat severity, (iii)~maintain service differentiation between malicious and legitimate users, and (iv)~automatically restore fair resource allocation after threat mitigation, demonstrating both defensive resilience and operational recovery.

\section{Conclusion and Future Work}\label{sec:future_work}
\greenub{CoMeT-Net addresses O-RAN anomaly detection challenges through three key innovations: structured memory banks that decompose detection into granular pattern matching, adaptive gating that downweights ambiguous features as a learned noise filter, and contrastive alignment that structures the latent space for template-based classification. By eliminating separate classifier networks through consensus voting, our approach achieves 99.35\% F1 score on network traffic with 10$\times$ lower false alarm rates and 0.3-3ms inference across hardware tiers from servers to Raspberry Pi 4. O-RAN testbed deployment validates practical effectiveness, isolating malicious traffic (RTT$>$1400ms) while preserving legitimate user QoS (15-20ms RTT). Cross-domain evaluation on financial data confirms architectural generalizability.}

\greenub{While CoMeT-Net achieves state-of-the-art performance, several directions warrant further investigation. \textbf{Multiclass attack classification} would enable granular threat identification beyond binary detection. This requires expanding memory templates to capture subtle inter-class differences (e.g., distinguishing DoS from reconnaissance attacks) and modifying the gating mechanism to handle multiple negative classes simultaneously, enabling targeted countermeasures based on attack type. \textbf{Adaptive memory updates} could address evolving threat landscapes where attackers modify tactics. We envision online template refinement through exponential moving averages of correctly classified samples, coupled with anomaly-triggered template expansion when novel attack patterns consistently evade existing templates. The challenge lies in balancing stability (preventing catastrophic forgetting) with adaptability (incorporating new threats). \textbf{Federated learning across O-RAN domains} would enable collaborative threat intelligence without sharing raw traffic data. Multiple operator networks could jointly train shared memory templates while preserving privacy, but this requires addressing statistical heterogeneity across domains and designing communication-efficient aggregation protocols compatible with O-RAN's distributed architecture. Finally, \textbf{CPU-optimized architectures} warrant exploration. While our ResNet backbone achieves excellent accuracy, its reliance on parallel computation limits throughput on CPU-only devices. Investigating lightweight backbones (e.g., MobileNet variants) or quantization techniques could enable broader deployment on cost-constrained edge infrastructure without sacrificing detection quality.}



\balance
\bibliographystyle{ieeetr}
\bibliography{ref.bib}

\end{document}